\documentstyle[12pt]{article}
\begin{document}
\large
\begin{center}

\null\vskip 5cm

EXTENDED COHERENT STATES AND MODIFIED PERTURBATION THEORY                                                                   

\vskip 5pt
                      G.M.Filippov

       Chuvash State University, Cheboksary, Russia

            E-mail: gennadiy@chuvsu.ru
\vskip 10mm
\end{center}

\begin{abstract}
  An extended coherent state (ECS) for describing a system of two 
interacting quantum objects is considered. A modified perturbation 
theory based on using the ECSs is formulated.
\end{abstract}

\vskip 10mm
\noindent
PACS numbers: 03.65.Fd, 11.15.Bt, 11.15.Tk
\vskip 5pt

\noindent
{\bf 1. Extended coherent states}
\vskip 3pt\noindent
Coherent states were constructed first by Schr\"odinger [1]
and in the last 40 years of the 20th century were widely used in
different problems of quantum physics [2]. There are many 
modifications of coherent states. Recall, for example, the spin coherent
states introduced in [3,4]. A general algebraic 
approach in the coherent state theory was developed in [4]. The 
coherent states for a particle on a sphere were applied in [5] to 
describe the rotator time evolution. Here we propose one more 
generalization of the theory by introducing the extended coherent 
state (ECS).

    Consider a system of an oscillator and a free spinless particle 
posessing a momentum ${\bf k}_0$. Let  ${\widehat b}^\dagger$ and 
${\widehat b}$  be the ladder operators for the oscillator. Introduce 
the creation ${\widehat a}^\dagger$ and annihilation ${\widehat a}$ 
operators of Bose type to describe a possible change in the particle's 
state (note that the further consideration may be applied just as well 
to a Fermi particle). Input the operator
\begin{equation}
{\widehat Q}=\sum\limits_{\bf q} h_{\bf q}{\widehat\rho}_{\bf q},
\end{equation}
where
$${\widehat\rho}_{\bf q}=\sum\limits_{\bf k}
{\widehat a}_{\bf k}^\dagger {\widehat a}_{\bf k+ q}$$
is the Fourier component of the density operator and $h_{\bf q}$ -
are coefficients depending on momentum $\bf q$. We can construct another 
linear combination ${\widehat Q}'$ of operators ${\widehat\rho}_
{\bf q}$ with the help of any other set of coefficients 
$h_{\bf q}'$. All these combinations are commutative
$$
[{\widehat Q},{\widehat Q}']_- = 0
$$
because the commutation rule
\begin{equation}\label{02}
[{\widehat\rho}_{\bf q},{\widehat\rho}_{\bf q'}]_- = 0
\end{equation}
is fullfilled for all $\bf q$ and $\bf q'$.

   Input a vector of state $| 0,{\bf k}_0)$, where the
first argument $(0)$ denotes a ground state of the oscillator
and the
 second one $({\bf k}_0)$ describes a state of the particle. 
Define the vector
\begin{equation}
| h,{\bf k}_0>=\exp{\left(-{1\over 2} {\widehat Q}^\dagger
{\widehat Q}\right)}\sum\limits_{n=0}^\infty
{1\over n!}({\widehat Q}{\widehat b}^\dagger)^n
| 0,{\bf k}_0)
\end{equation}
as an ESC (here we briefly denote by $h$ the whole set of coefficients 
$h_{\bf q}$). Obviously, the vector (3) coincides with the ordinary 
Schr\"odinger coherent state (SCS), when one replaces all particle's 
operators by their classical equivalents. The ECS describes some 
state of a system of two interacting quantum objects --- the particle and 
the oscillator. By this circumstance the ECS sufficiently differs from 
the SCS.

   We outline the following general properties of the ECS:
\par\noindent
(1) The ECS is not the eigenvector for ${\widehat b}$, but
\begin{equation}
{\widehat b}\:|h,{\bf k}_0>= {\widehat Q} |h ,{\bf k_0}>. 
\end{equation}
\par\noindent\hangindent 20pt
(2) The operators ${\widehat\rho}_{\bf q}$ only change  momenta  for
all the one-particle states. Hence, the following relations are 
fullfilled:
\begin{equation}
{\widehat\rho}_{\bf q} | h,{\bf k}_0>=
| h,{\bf k}_0-{\bf q}>
\end{equation}
\begin{equation}
{\widehat\rho}_{\bf q}^\dagger {\widehat\rho}_{\bf q} 
| h,{\bf k}_0>= | h,{\bf k}_0>.
\end{equation}
\par\noindent\hangindent 20pt
(3) There is the following representation:
\begin{equation}
| h,{\bf k}_0>=\exp [{\widehat Q}
{\widehat b}^\dagger- {\widehat Q}^\dagger 
{\widehat b}] |0,{\bf k}_0)
\end{equation}
which is equivalent to the relevant representation of the SCS.
\par\noindent\hangindent 20pt
(4) If $h_{\bf q} = g\,\Delta ({\bf q}-{\bf q}_0)$ we easily have 
\begin{equation}
| h,{\bf k}_0>=\exp \left[{- | g |^2\over 2}\right]
\sum\limits_{n=0}^\infty {g^n \over n!} ({\widehat \rho}_{{\bf q}_0}
{\widehat b}^\dagger)^n | 0,{\bf k}_0)
\end{equation}
and therefore,
\begin{equation}
< h,{\bf k}_0\vert h',{\bf k}_0'> = \exp\biggl[-
{1\over 2}( |g|^2+| g'|^2-2 g^*
g') \biggr] \Delta({\bf k}_0-{\bf k}'_0).
\end{equation}
\par\noindent\hangindent 20pt
(5) The total amount of ECS are more than sufficient to define the
Hilbert space. Following to Klauder [6] (see also [7])
we can introduce the development of the unity operator
\begin{equation}
{\widehat {\rm I}}=\sum\limits_{\bf k}{1\over \pi}\int d^2 z
\,{\widehat Q}\; | z h, {\bf k}>< z h, {\bf k}| 
\;{\widehat Q}^\dagger
\end{equation}
where $z$ - is the complex variable,  $d^2 z=d[{\rm Re}(z)]
 d[{\rm Im}(z)]$. To prove the last equation one may use the
integral
$$
\int d^2 z \,(z^*)^n z^m\: \exp \biggl[-|z|^2 {\widehat Q}^\dagger
{\widehat Q}\biggr]\; {\widehat Q}^{m+1} 
({\widehat Q}^\dagger)^{n+1} = \pi\, n!\: \delta_{nm}.
$$
\par\noindent\hangindent 20pt
(6) There is the following useful sum rule:
\begin{equation}
\sum\limits_{\bf k} e^{i{\bf s}{\bf k}} {\widehat a}_
{\bf k} | h,{\bf k_0}>=e^{i{\bf s k}_0}
| \alpha)\otimes | {\rm vac}_p)
\end{equation}
where the right-hand side contains a direct product of the SCS for the 
oscillator 
$$
| \alpha)=\exp \left[-{1\over 2} | \alpha |^2 \right]
\sum\limits_0^\infty {\alpha^n\over n!}(b^\dagger)^n | 0)
$$ 
and a vacuum state of the particle $ |{\rm vac}_p)$. Here 
the quantity $\alpha$ is given by the formula
$$
\alpha = \sum_{\bf q} h_{\bf q} e^{-i s {\bf q}}.
$$
To prove the property (6) one should keep in mind the relation:
\begin{equation}
\sum\limits_{\bf k} {\widehat a}_{\bf k} e^{i\bf kx} 
{\widehat\rho}_{{\bf q}_1} {\widehat\rho}_{{\bf q}_2}...
|0,{\bf k}_0)=e^{i{\bf k}_0{\bf x}}
e^{-i{\bf q}_1{\bf x}} e^{-i{\bf q}_2{\bf x}}
...|0)\otimes |vac_p)
\end{equation}
where $|0)$ is the vector of the ground state of the oscillator.

\vskip 10pt
\noindent
{\bf   2. Modified perturbation theory}
\vskip 3mm

\noindent
   ECSs, first introduced in 1983[8]\footnote
{ Extended coherent states were first
denoted as 'double coherent' states or 'modified coherent' states.}
arise, for example, in a problem of interaction between a moving 
particle and an oscillator. The proper Hamiltonian can be represented 
in the following general form
\begin{equation}\
{\widehat H}_{int}={\widehat b}^\dagger \sum\limits_{\bf q}
g_{\bf q} {\widehat \rho}_{\bf q}+ {\widehat b}\sum\limits_{\bf q}
g_{\bf q}^* {\widehat \rho}^\dagger_{\bf q}
\end{equation}
where $g_{\bf q}$ - is a coupling function. Since ${\widehat\rho}_
{\bf q}^\dagger = {\widehat\rho}_{-\bf q}$, it should be $g_{-\bf q}=
g_{\bf q}^*$.

   In most applications the Hamiltonian (13) within the interaction 
picture depends on time via the density operators ${\widehat\rho}(t)$.
In these cases we cann't apply ECS without some modification of the
theory.  Indeed, instead of relations (2) we have
$$
[{\widehat\rho}_{\bf q}(t),{\widehat\rho}_{\bf q'}(t')]_- = 
\sum\limits_{\bf k} {\widehat a}^\dagger_{\bf k} {\widehat a}_{\bf k+q+
q'} \Bigl[ \exp\{i(\varepsilon_{\bf k}t-\varepsilon_{\bf k+q+
q'}t'-i\varepsilon_{\bf k+q}(t-t')\}
$$
$$
- \exp\{i(\varepsilon_{\bf k}t'-\varepsilon_{\bf k+q+q'}t+
i\varepsilon_{\bf k+q'}(t-t')\}\Bigr].
$$
We construct a modified perturbation theory with the help of excluding 
an integrable part of the interaction. For this purpose we expand the 
operator ${\widehat H}_{int}(t)$ in two parts,  
${\widehat H}^{(0)}_{int}(t)$ and ${\widehat H}^{(1)}_{int}(t)$,
where
$$
{\widehat H}^{(0)}_{int}(t)={\widehat b}^\dagger \sum\limits_{\bf q}
g_{\bf q} {\widehat \rho}_{\bf q} f_{\bf q}(t)+ 
{\widehat b}\sum\limits_{\bf q} g_{\bf q}^* {\widehat \rho}^\dagger_{\bf q}
f_{\bf q}^*(t)
$$
$$
{\widehat H}^{(1)}_{int}(t)={\widehat H}_{int}(t)-{\widehat H}^{(0)}
_{int}(t).
$$
Here the function $f_{\bf q}(t)$ must be unimodular to preserve
the interaction intensity. Obviously, the operators 
$\sum\limits_{\bf q} g_{\bf q} {\widehat \rho}_{\bf q} f_{\bf q}(t) $
defined at different times, obey the commutation relations. 
Then, by virtue of the above consideration, the equation
$$ 
i{d\over dt} | t)={\widehat H}_{int}^{(0)}(t) | t)
$$
acquires an exact solution
\begin{equation}
| t)=e^{-i{\widehat\chi}(t)} |h,{\bf k}_0>
\end{equation}
where ${\widehat Q}$ has the previous form (1) and
$$
h_{\bf q}=-i g_{\bf q} \int\limits_0^t dt' 
f_{\bf q}(t')\: e^{i\omega t'}
$$
$$
{\widehat\chi}(t)=-{i\over 2}\int\limits_
0^t\lbrace{\widehat{\dot Q}}^\dagger(t')
{\widehat Q}(t')-{\widehat Q}^\dagger (t')
{\widehat{\dot Q}}(t')\rbrace dt'.
$$

The solution (14) can be rewritten as $| t)={\widehat U}_0(t)
| 0,{\bf k}_0>$, where we introduce a zero-th order evolution 
operator
$${\widehat U}_0(t)=\exp \lbrace
{\widehat Q}(t) {\widehat b}^\dagger
-{\widehat Q}^\dagger (t) {\widehat b}
-i {\widehat{\chi}}(t)\rbrace.$$
There are the following useful commutation relations:
$$
[{\widehat b},{\widehat U}_0(t)]_{-}={\widehat U}_0(t){\widehat Q}(t)
\qquad\quad [{\widehat b},{\widehat U}_0^\dagger(t)]_{-}=-
{\widehat U}_0^\dagger (t){\widehat Q}(t)
$$
$$
[{\widehat b}^\dagger,{\widehat U}_0(t)]_{-}={\widehat U}_0(t)
{\widehat Q}^\dagger(t)
\qquad\quad [{\widehat b}^\dagger,{\widehat U}_0^\dagger(t)]_{-}=-
{\widehat U}_0^\dagger(t){\widehat Q}^\dagger (t).
$$

Let us introduce a new representation for the vector of state and
for operators:
$$
|t>={\widehat U}^\dagger_0(t)|t) \qquad\quad {\widetilde A}={\widehat  U}_0^\dagger
(t) {\widehat A}{\widehat U}_0(t).
$$
The new vector of state obeys the equation
$$
i {d\over dt}|t>= {\widetilde H}^{(1)}_{int} (t) |t>
$$
which can be solved with the help of a standard technique using the 
T-exponent
\begin{equation}
|t>={\rm Texp}\{-i\int\limits_0^t dt' {\widetilde H}_{int}^{(1)}(t')\}\;
|0,{\bf k}_0).
\end{equation}

If the choice of the function $f_{\bf q}(t)$ ensures the rapid 
convergence to the series (15), formula (14) gives a good 
approximation for the vector of state. In this case we can evaluate 
a wide set of physical characteristics with sufficient accuracy. 
As an example, we calculate the density matrix for the  particle,  for
which the exact expression is given by the formula
\begin{equation}
\Gamma({\bf x},\,{\bf x}',\,t)=<t|{\widetilde \psi}^\dagger
({\bf x},t){\widetilde\psi}({\bf x}',t)|t>.
\end{equation}
Here the usual wave operators are introduced, namely,
$$
{\widetilde\psi}({\bf x},t)={\widehat U}_0^\dagger(t) {\widehat\psi}
({\bf x},t) {\widehat U}_0 (t) \quad\qquad
{\widehat\psi}({\bf  x},t)=\sum\limits_{\bf  k}{\widehat  a}_{\bf   k}
\exp\{i{\bf k x}-i\varepsilon_{\bf k}t\}
$$
where $\varepsilon_{\bf k}$ - is an energy of the particle posessing 
momentum ${\bf k}$.
The further consideration will be more convenient if the particle-
oscillator interaction began at any incident time  $t_0<0$  when  the
oscillator was found in the ground  state.  Let us  define  the  density
matrix at $t=0$. In the  first  approximation  we  can  set  $|t>\approx
|0,{\bf k}_0)$.  In this case
\begin{equation}
\Gamma({\bf x},\,{\bf x}',\,t)\approx (0,{\bf k}_0|{\widehat U}_0^\dagger
(0){\widehat\psi}^\dagger ({\bf x},0) {\widehat\psi}({\bf x}',0)
{\widehat U}_0 (0)|0,{\bf k}_0).
\end{equation}
Using relation (7) we have ${\widehat U}_0 (0) |0, {\bf k}_0) = 
e^{-i{\widehat \chi} (0)} |h, {\bf k}_0>$, where ${\widehat Q}$ 
is defined as in (1) with
$$
h_{\bf q}=h_{\bf  q}(0) \qquad
h_{\bf q}(t)=-i g_{\bf q}\int\limits_{t_0}^t  f_{\bf  q}(t')
e^{i\omega  t'} dt'\qquad t>t_0.
$$
Now we apply relations (12) to obtain the formula similar to (11):
\begin{equation}
{\widehat\psi}({\bf x}', 0) {\widehat U}_0 (t) |0,{\bf k}_0) =
\exp\{i{\bf k}_0{\bf x}' - i\Phi({\bf x'})\} 
|\alpha({\bf x'},0))\otimes |vac_p) 
\end{equation}
where
$$
\alpha({\bf x}, t) = \sum\limits_{\bf q} h_{\bf q} (t) e^{-i{\bf qx}}
$$
$$
\Phi({\bf x}) = \int\limits_{t_0}^0 {\rm Im}\left[ {\dot\alpha}^*({\bf x},t')
\alpha({\bf x},t')\right]\,dt'.
$$
Substituting (18) into (17) we obtain
\begin{equation}
\Gamma({\bf x},\,{\bf x}',\,0)\approx e^{-i{\bf k}_0{\bf x}+i{\bf k}_0
{\bf x}'} \times
$$
$$
\exp\biggl\{i\Phi({\bf x})-i\Phi({\bf x}')-{1\over 2}\biggr[
|\alpha({\bf x},0)|^2 + |\alpha({\bf x}',0)|^2 - 2\alpha^* ({\bf x},0)
\alpha({\bf x}',0)\biggr]\biggr\}.
\end{equation}
Note, that in the case  $g_{\bf q} = g \Delta ({\bf q}-{\bf q}_0)$, the
phase  $\Phi({\bf x}) = const$ and formula (19) is simlified.

The work was partly supported by the Russian Foundation for Basic 
Research (grant no 97-02-16058).

\end{document}